\documentstyle[epsf]{ptptex}

\title{Wave Propagation in Nonlocally Coupled Oscillators with Noise }

\author{Yuri {\sc Shiogai} and Yoshiki {\sc Kuramoto}}  

\inst{Department of Physics, Graduate School of Sciences, 
      Kyoto University, Kyoto 606-8502,Japan}

\recdate{ }

\abst{The onset of undamped wave propagation in noisy 
self-oscillatory
media is identified with a Hopf bifurcation of the corresponding 
effective dynamical system obtained by properly renormalizing 
the effects of noise. We illustrate this fact on
a dense array of nonlocally coupled 
phase oscillators  
for which a mean-field idea works exactly in deriving such effective
dynamical equations.}

\begin{document}

\maketitle

Unlike conservative oscillatory media, dissipative self-oscillatory
media admit traveling waves without decay even in the presence of noise or
other sources of randomness. This ability of wave transmission in 
random self-oscillatory
media should be functionally relevant in a variety of living 
organisms for which randomness is unavoidable. For this type of systems, 
a critical strength of randomness is generally expected to exist such that
below which the system is capable of sustaining undamped traveling waves. 
This critical point should be identical with the 
Hopf bifurcation point of an effective dynamical system obtained by properly
renormalizing the effects of noise. It is reasonable to expect that 
a theory could be developed unambiguously on this issue in the particular case
when each oscillator couples with sufficiently many
oscillators, because a mean-field idea should be applicable then.
In the present article, we carry out this program for nonlocally coupled phase
oscillators with noise, and show how an effective dynamical equation
can be derived, and how it is reduced to
a small-amplitude equation near the bifurcation point.
Our theory may be regarded
as a natural extension of a previous theory\cite{rf:1} 
on the onset of collective 
oscillation for globally coupled phase oscillators with noise.

Imagine an infinitely long array of nonlocally coupled oscillators which are
distributed densely and subject to additive noise. By taking the continuum limit, the phases  
$\phi(x,t)$ of the oscillators are assumed to obey the following 
Langevin-type equation.
\begin{equation}
\dot{\phi}(x,t) = \omega + \int_{-\infty}^{\infty}G(x-x')\Gamma(\phi(x,t)-\phi(x',t'))+\xi(x,t). 
\end{equation}
Here $\omega$ is the natural frequency, the second term represents the nonlocal
coupling, and the last term gives additive noise. The phase coupling function 
$\Gamma$
depends only on the phase difference and satisfies the in-phase condition
$\Gamma'(0)<0$\cite{rf:1}
, while
its strength $G$ depends on the distance. Specifically, we assume
the form 
$G(x)={\gamma}\exp(-{\gamma}|x|)/2$ whose integral is normalized.
The noise is assumed to be white Gaussian with vanishing mean, i.e.,
$<\xi(x,t)>=0$ and $<\xi(x,t)\xi(x',t')>=2D\delta(t-t')\delta(x-x')$.
Since we are working with an oscillator continuum, 
the coupling radius $\gamma^{-1}$ 
is so large that
infinitely many local oscillators may fall within it. 
This allows us to apply a mean field
idea similar to the one applied successfully to globally coupled oscillators
involving randomness.
We also remark that the model given by Eq.\ (1) is not merely our invention
but the one derived under suitable conditions from a certain 
class of reaction-diffusion systems after eliminating adiabatically a highly
diffusive chemical component\cite{rf:2}.

Periodic traveling waves in our oscillatory field correspond to a family of solutions of Eq.\ (1) having the form ${\phi}=kx+{\Omega}t$ possibly disturbed
by the noise. In the absence
of noise, it is easy to confirm their existence 
and analyze their stability for general phase-coupling function.
In the simple case given by
\begin{equation}
 \Gamma(\phi)=-\sin(\phi+\alpha) \quad (|\alpha|<\pi/2),
\end{equation}
for instance,
the traveling waves of sufficiently small $k$ turn out always stable.
The additive noise is now switched on. Numerical simulation for 
$|\alpha|$ not too close to $\pi/2$ shows that as
the noise becomes stronger, traveling waves of shorter wavelengths (still
satisfying the stability condition for the noiseless case) 
successively become unstable. For sufficiently strong noise, no traveling 
waves can exist, implying that the 
medium is no longer self-oscillatory
because the traveling wave with infinite wavelength corresponds
to a uniform oscillation.  
In order to explain these results, we will now convert Eq.\ (1) to a 
nonlinear
Fokker-Planck equation and apply the center-manifold reduction to it.

In terms of the number density $n(\psi,x,t)\equiv \delta(\phi(x,t)-\psi)$,
Eq.\ (1) can be rewritten as
\begin{eqnarray}
\frac{\partial\phi(x,t)}{\partial t}&=&\omega+\int_{-\infty}^{\infty}
dx'G(x-x')\int_{0}^{2\pi}d\psi'\Gamma(\phi(x,t)-\psi')n(\psi',x',t)
+\xi(x,t). \nonumber\\
&\equiv& V(\phi(x,t),x,t)+\xi(x,t).
\end{eqnarray}
The drift velocity $V$ introduced above is expressed as
\begin{equation}
V(\psi,x,t)=\omega+\int_{0}^{2\pi}d\psi'Z(\psi',x,t)\Gamma(\psi-\psi')d\psi',
\end{equation}
where
\begin{equation}
Z(\psi',x,t)=\int_{-\infty}^{\infty}dx'G(x-x')n(\psi',x',t).
\end{equation}
The quantity $Z$ may be interpreted as a space-time dependent mean field.
If we regard $Z$ as an externally given quantity, Eq.\ (3)
represents a non-autonomous one-oscillator equation. Thus, the normalized
probability density
$f(\psi,x,t)$ such that $\phi(x,t)$ takes value $\psi$ is governed by the
Fokker-Planck equation\cite{rf:3}
\begin{equation}
\frac{\partial f(\psi,x,t)}{\partial t}=-\frac{\partial}{\partial\psi}\bigl[
V(\psi,x,t)f(\psi,x,t) \bigr]+D\frac{\partial^{2}f(\psi,x,t)}{\partial\psi^2}.
\end{equation}
We now take a statistical average of the 
dynamical variable $n(\psi',x',t)$ appearing in the mean field $Z$. 
This is allowed because $Z$ is given by a weighted sum over infinitely many
oscillators which are individually subject to random forcing so that
the statistical fluctuation of $Z$ may safely be neglected.  
Noting that the average $<n(\psi',x',t)>$ is identical
with $f(\psi',x',t)$ by definition, we finally find a nonlinear evolution 
equation for
$f$ in the form
\begin{eqnarray}
\frac{\partial f(\psi,x,t)}{\partial t}&=&-\frac{\partial}{\partial \psi}
\bigl[\bigl\{\omega+\int_{-\infty}^{\infty}dx'G(x-x')\int_{0}^{2\pi}d\psi'
\Gamma(\psi-\psi')f(\psi',x',t)\bigr\}f(\psi,x,t) \bigr] \nonumber\\
&&+D\frac{\partial^{2}f(\psi,x,t)}{\partial\psi^2}.
\end{eqnarray}
Wave propagation and collective dynamics could be discussed on the basis of
this deterministic kinetic equation. In particular, as is shown below,
one may apply the
center-manifold reduction near the critical noise strength $D=D_c$ at which
the trivial constant solution $f_{0}=(2\pi)^{-1}$ loses stability.
For this purpose, it is convenient to rewrite Eq.\ (7) in terms of the 
fluctuation $\rho(\psi,x,t)$ defined by $f(\psi,x,t)=f_0+\rho(\psi,x,t)$.
The $2\pi$-periodic functions $\rho(\psi)$ and $\Gamma(\psi)$ are further 
developed into Fourier series
like $\rho(\psi,x,t) = {2\pi}^{-1}\sum_{l\neq0}\rho_l(x,t)\exp(il\psi)\nonumber$ and $\Gamma(\psi)=\sum_l\Gamma_l\exp(il\psi)$. Using these expressions,
Eq.\ (7) becomes
\begin{eqnarray}
\frac{\partial\rho_{l}(x,t)}{\partial t}&=&-[l^{2}D+il(\omega+\Gamma_{0})]
\rho_{l}(x,t)-il\Gamma_{l}\int_{-\infty}^{\infty}dx'G(x-x')\rho_{l}(x',t)
\nonumber\\
&&-il\int_{-\infty}^{\infty}dx'G(x-x')\sum_{m\ne 0,l}\Gamma_{m}\rho_{m}(x,t)
\rho_{l-m}(x,t).
\end{eqnarray}
Regarding the loss of stability of the constant solution $f_{0}$, 
spatially uniform perturbation is the most dangerous. This is because
by definition the uniform Fourier component of $G$ is the largest of all its 
Fourier components, thus 
making the linear growth rate of the uniform perturbation
of $\rho_{l}$ the largest.
Therefore, the sign of this growth rate $\sigma_l$ defined by
\begin{equation}
\sigma_{l}=-l^2 D+l\Im(\Gamma_{l})
\end{equation} 
determines the stability of $f_{0}$ with respect to the fluctuation $\rho_l$.
Note that $\Im(\Gamma_{-l})=-\Im(\Gamma_{l})$, so that $\sigma_l$ is 
independent of the sign of $l$.
Suppose that $\sigma_{l}$ as a function of $l$ takes the largest value
at $l=\pm\lambda$ $(\lambda>0)$. 
As the noise becomes weaker, a Hopf bifurcation occurs
 at $D=D_{c}=\Im(\Gamma_{\lambda})/\lambda$, the corresponding frequency 
being given by $\Omega=
 -i\lambda(\omega+\Re(\Gamma_{\lambda})+\Gamma_{0})$.

The center-manifold reduction can be achieved almost in parallel with 
the case of globally coupled phase oscillators with noise\cite{rf:1}.
The only difference is the existence of spatial integrals
in the governing equation. Since the
characteristic spatial scale is expected to be very large near the bifurcation
point, which can actually be confirmed from the results obtained,
one may develop $\rho(x')$ into a Taylor series like
$\rho(x')=\rho(x)+(x'-x)\rho'(x)+(x'-x)^{2}\rho''(x)/2+\cdots$ in Eq.\ (8),
and neglect higher order effects. It turns out that the replacement
$\rho(x')$ with $\rho(x)$ is allowed in the nonlinear terms, while
the second order derivative of $\rho(x)$ must be included in the linear term.
In this way, Eq.\ (8) takes the form of a reaction-diffusion system,
and its reduced form near $D_c$ is the complex Ginzburg-Laudau equation.
As usual, a complex amplitude $A(x,t)$ related to $\rho(\psi,x,t)$ is
introduced as
\begin{equation}
\rho(\psi,x,t)=\frac{1}{2\pi}\bigl(A(x,t)\exp[i(\lambda\psi+\Omega t)]
+c.c.\bigr).
 \end{equation}
By applying the standard method, the reduced equation near $D=D_{c}$ 
becomes
\begin{equation}
\frac{\partial A}{\partial t}=\lambda^{2}(D_{c}-D)A
+d
\frac{\partial^{2}A}{\partial x^{2}}
-g|A|^{2}A,
\end{equation}
where $d$ and $g$ are complex coefficients given by
\begin{equation}
d=\frac{i\lambda\Gamma_{\lambda}}{\gamma^{2}}, \quad
g=\frac{\lambda\Gamma_{\lambda}(\Gamma_{2\lambda}+\Gamma_{-\lambda})}
{2\Im(\Gamma_{\lambda})-i\Re(\Gamma_{\lambda})+i\Gamma_{2\lambda}}.
\end{equation}

We now consider the simple coupling function given by Eq.\ (2), and
show some results for this particular case.
Obviously, we have $\lambda=1$, 
$\Gamma_{\pm 1}=\pm i\exp(i\alpha)/2$ and
all $\Gamma_{l}$ with $|l|\ge 2$ vanish. The quantities 
$d$ and $g$ are simplified as
\begin{eqnarray}
d&=&-\frac{i\Gamma_{1}}{\gamma^{2}}=\frac{1}{2\gamma^{2}}\exp(i\alpha),
\\
g&=&\frac{|\Gamma_{1}|^{2}}{2\Im(\Gamma_{1})-i\Re(\Gamma_{1})}=\frac{1}{4\cos
\alpha+2i\sin\alpha}.
\end{eqnarray}
In this particular case, the bifurcation is supercritical because $\Re(g)>0$.

From the well-known properties of the complex Ginzburg-Landau equation,
Eq.\ (11) admits a family of traveling plane waves $A_{k}(x,t)$ in the form
$A_{k}=R_{k}\exp[i(kx+\Omega_{k}t)]$. The stability condition of the uniform 
solution $A_{0}$ is given by $1+c_{1}c_{2}>0$, where $c_{1}=\Im(d)/\Re(d)$ and
$c_{2}=\Im(g)/\Re(g)$. For the sine-coupling given by Eq.\ (2),
the corresponding condition becomes $\tan^{2}\alpha<2$ or $|\alpha|<0.9553...$.
Under this condition, plane waves with $k$ up to some critical value
of order $(D_c-D)^{1/2}$ can propagate through the medium stably. This is
consistent with the aforementioned results of our
numerical simulation on the stochastic system Eq.\ (1).

Note that 
if $|\alpha|$ exceeds the above critical 
value, the uniform oscillation becomes unstable and phase turbulence sets in.
In terms of the original model given Eq.\ (1), this means that
the noise can strongly be amplified to a turbulent level.  Note also
that Eq.\ (11) admits the Nozaki-Bekki hole solutions\cite{rf:4} 
for which
the existence of amplitude degree of freedom is crucial. If the phase model 
Eq.\ (1) which apparently lacks the amplitude variables can also
exhibit hole structures, this should solely 
be due to a strong phase scattering
caused by the noise in a local group of phase oscillators
by which the effective amplitude there becomes vanishingly small.
Detailed numerical study associated with these remarks will be reported
elsewhere.

\end{document}